\documentclass[
    ,final            
  ]
  {aipproc}

\layoutstyle{6x9}

\begin{document}
\newcommand{\as}{\alpha_s}
\newcommand{\be}{\begin{eqnarray}}
\newcommand{\ee}{\end{eqnarray}}
\newcommand{\ben}{\begin{eqnarray*}}
\newcommand{\een}{\end{eqnarray*}}
\def\eq#1{{Eq.~(\ref{#1})}}
\def\fig#1{{Fig.~\ref{#1}}}
\newcommand{\un}[1]{\underline{#1}}
\newcommand{\stackeven}[2]{{{}_{\displaystyle{#1}}\atop\displaystyle{#2}}}
\newcommand{\lsim}{\stackeven{<}{\sim}}
\newcommand{\gsim}{\stackeven{>}{\sim}}

\title{Saturation Physics in Heavy Ion Collisions}

\author{Yuri V. Kovchegov}{
  address={Department of Physics, University of Washington, Box
  351560, Seattle, WA 98195}
}

\begin{abstract}
We discuss expectations of saturation physics for various observables
in heavy ion collisions. We show how simple saturation-inspired
assumptions about particle production in heavy ion collisions lead to
Kharzeev-Levin-Nardi model. Comparing this model to RHIC data on
particle multiplicities we conclude that saturation effects may play
an important role in particle production and dynamics at the early
stages of $Au-Au$ collisions already at RHIC energies. We then
estimate the contribution of the initial state two-particle azimuthal
correlations to elliptic flow observable $v_2$ in $Au-Au$ collisions
by constructing a lower bound on these non-flow effects based on $v_2$
obtained from the analysis of proton-proton ($pp$) collisions.
\end{abstract}

\maketitle


\section{Introduction}

Saturation/Color Glass Condensate physics is based on the observation
that the small-$x$ wave functions of ultrarelativistic hadrons and
nuclei are characterized by a hard scale $Q_s$, known as the
saturation scale \cite{glr,mq,BM,mv}. The scale $Q_s$ arises due to
{\sl saturation} of partonic densities at small-$x$ and is an
increasing function of energy and atomic number of the nucleus
\cite{bk,LT,mv}. This large scale makes the strong coupling constant small 
$\alpha_s (Q_s) \ll 1$ leading to dominance of the classical gluonic
fields in all high energy processes \cite{mv,kjklw}. Gluon production
in high energy collisions is given by the classical field of the
scattering color charges \cite{kmw}. Corresponding gluon production
cross section was found for $pA$ collisions in \cite{KM} and the
effects of quantum evolution \cite{bk} were included in it in
\cite{KT}. The gluon production cross section for heavy ion collisions
($AA$) at the classical level has been studied both numerically
\cite{KV} and analytically \cite{yuriaa}. Since it is quite not clear
at present how to include the effects of nonlinear quantum evolution
\cite{bk} in the results of \cite{KV,yuriaa}, one has to construct
models to describe the actual rapidity-dependent data produced in
heavy ion collisions. Below we are going to show how some of these
models, based on rather basic properties of saturation physics,
provide a reasonably good description of RHIC data.

\section{Particle Multiplicity from Saturation Models}

\subsection{Multiplicity at Mid-Rapidity Versus Centrality}

Classical field $A_\mu \sim 1/g$ leads to produced gluon multiplicity
\be
\frac{d N}{d^2 k \, d^2 b \, dy} \, \sim \, \left< A_\mu A_\mu \right> 
\, \sim \, \frac{1}{\as}.
\ee
Gluon transverse momentum spectrum described by a single scale $Q_s$
can be written as
\be
\frac{d N}{d^2 k \, d^2 b \, dy} \, = \, \frac{1}{\as} \, f(k_\perp / Q_s)
\ee
with some unknown function $f(k_\perp / Q_s)$ to be determined by
actual calculations. Integrating over $k$ and $b$ yields
\be\label{dndy}
\frac{d N}{dy} \, = \, \mbox{const} \, \frac{1}{\as}
\, \pi \, R^2 \, Q_s^2.
\ee
where the value of the constant is determined from $f(k_\perp /
Q_s)$. Following \cite{KN,KL} we assume that the scale for the
coupling constant in \eq{dndy} is set by $Q_s$. (This step, of course,
goes beyond the classical limit and assumes that \eq{dndy} is valid
even when running coupling corrections are included.) Then, as $R^2 \,
\sim \, A^{2/3}$ and if $Q_s^2 \, \sim \, A^{1/3}$ \cite{mv,kjklw}, 
together with running coupling
\be
\as (Q_s) \, = \, \frac{1}{b \, \ln Q_s^2/\Lambda^2} \sim \,  \frac{1}{\ln A}
\ee
we conclude from \eq{dndy} that
\be\label{adep}
\frac{1}{A} \, \frac{d N}{d\eta} \, \sim \, \ln A.
\ee
For heavy ion experiments at different collision centralities we
substitute $A$ by the number of participants $N_{part}$ so that
\eq{adep} becomes
\be\label{ndep}
\frac{1}{N_{part}} \, \frac{d N}{d\eta} \, \sim \, \ln N_{part}.
\ee
\begin{figure}[b]
  \includegraphics[height=.22\textheight]{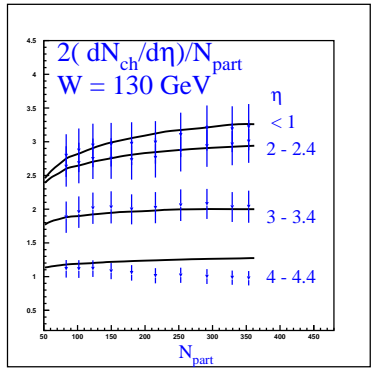}
  \caption{Saturation model fit of the PHOBOS data on total charged
  particle multiplicity at mid-rapidity as a function of centrality at
  $\sqrt{s} = 130$~GeV taken from \cite{KL}.}
\label{dnpart}
\end{figure}
\eq{ndep} allowed the authors of \cite{KN} to correctly {\sl predict} the 
particle multiplicity at mid-rapidity at RHIC as a function of
centrality at $\sqrt{s} = 130$~GeV. A fit of particle multiplicity at
other values of rapidity at $\sqrt{s} = 130$~GeV taken from \cite{KL}
is shown in \fig{dnpart}.

\subsection{Multiplicity as a Function of Energy}

\eq{dndy} allows one to test whether the scaling of total particle 
multiplicity with energy is consistent with saturation/Color Glass
predictions. Using the fact that $Q_s^2 \sim 1/x_{Bj}^\lambda$
\cite{bk,LT} in \eq{dndy} and, for the moment, dropping the slower 
$Q_s$-dependence in $\as$ leads to
\be
\frac{dN/d\eta (\sqrt{s_1})}{dN/d\eta (\sqrt{s_2})} \, = \,
\left( \frac{\sqrt{s_1}}{\sqrt{s_2}} \right)^{\lambda}.
\ee
Using PHOBOS data for total charge multiplicity at $\sqrt{s} =
130$~GeV for most central collisions
\be
\frac{d N}{d\eta} (\sqrt{s} = 130 \, \mbox{GeV}) \, = \, 555 \pm 12 (stat) 
\pm 35 
(syst)
\ee
together with $\lambda = 0.25 \div 0.3$ obtained in \cite{GBW} by
analyzing HERA data, Kharzeev and Levin \cite{KL} {\sl predicted} the
total charge multiplicity at $\sqrt{s} = 200$~GeV to be
\be
\frac{d N}{d\eta} (\sqrt{s} = 200 \, \mbox{GeV}) \, = \, 616 \div 634
\ee
in agreement with the later measured PHOBOS result
\be
\frac{d N}{d\eta} (\sqrt{s} = 200 \, \mbox{GeV}) \, = \, 650 \pm 35 (stat).
\ee

\subsection{$\frac{d N}{d \eta}$ Versus $\eta$ and $N_{part}$}

Describing rapidity distribution of the produced particles requires a
little more modeling. Assuming $k_T$-factorization for particle
production cross section
\begin{figure}[b]
  \includegraphics[height=.23\textheight]{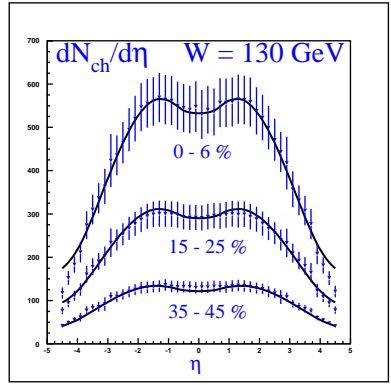}
  \caption{Saturation model fit of the PHOBOS data on total charged
  particle multiplicity as a function of rapidity and centrality at
  $\sqrt{s} = 130$~GeV taken from \cite{KL}.}
\label{deta}
\end{figure}
\be\label{kt}
\frac{d \sigma^{AA}}{d^2 k \ dy} \, = \, \frac{2 \, \as}{C_F} \, 
\frac{1}{{\un k}^2} \, \int d^2 q \, \phi_A ({\un q}) \, \phi_A ({\un k} 
- {\un q})
\ee
along with saturation-inspired unintegrated gluon distribution
functions ($\phi_A ({\un k}) \sim \as/{\un k}^2$ if $k_\perp > Q_s$
and $\phi_A ({\un k}) \sim S_\perp /\as$ if $k_\perp < Q_s$) the
authors of \cite{KL} produced an impressive fit of the PHOBOS data on
charged particle multiplicities as functions of rapidity and
centrality at $\sqrt{s} = 130$~GeV shown in \fig{deta}. The {\sl
predictions} made in \cite{KL} for similar multiplicity data at
$\sqrt{s} = 200$~GeV were also in good agreement with the later
published BRAHMS data.

Phenomenological success of the saturation models presented above does
not contradict the possibility of strong final state interactions
leading to formation of quark-gluon plasma. As was argued in
\cite{BMSS}, thermalization in the saturation framework would not introduce 
any fundamentally new scales leaving Eqs. (\ref{dndy}) and
(\ref{ndep}) practically unchanged. Late stage interactions are also
not very likely to significantly modify the rapidity distribution of
\fig{deta} due to causality constraints.

\section{Non-Flow Contribution to $v_2$}

The contribution of non-flow two-particle azimuthal correlations from
the early stages of heavy ion collisions to the elliptic flow
observable $v_2$ has been estimated in \cite{KTf} using a
saturation-inspired model of particle correlations. Here we are going
to construct a model-independent lower bound on these non-flow effects
using $v_2$ extracted from the analysis of $pp$ data. We start with
the definition of $v_2 (p_T)$ for $pp$ corrected for uncertainty in
the ``reaction plane'' definition (or, equivalently, defined through
the two-particle correlation functions, such that $v_2 (p_T) < v_2 >
\, = \, <\cos 2 (\phi_p - \phi_k)>_k$ )
\be\label{v21}
v_2^{pp} (p_T) \, = \, \frac{\int d^2 k \, \frac{d N_{corr}^{pp}}{d^3
p \, d^3 k} \, \cos 2 (\phi_p - \phi_k)}{\frac{d N^{pp}}{d^3 p} \,
\frac{d N^{pp}}{d y_k} + \int d^2 k \, \frac{d N_{corr}^{pp}}{d^3
p \, d^3 k}} \, \sqrt{\frac{\frac{d N^{pp}}{d y_p} \, \frac{d
N^{pp}}{d y_k} + \frac{d N_{corr}^{pp}}{d y_p \, d y_k}}{\int d^2 p \,
d^2 k \frac{d N_{corr}^{pp}}{d^3 p \, d^3 k} \, \cos 2 (\phi_p -
\phi_k)}}.
\ee
If we assume that the relative magnitude of the correlated terms in
\eq{v21} compared to uncorrelated ones is roughly the same for all $p_T$, 
we can drop the former compared to the latter finding that \eq{v21} is
approximately bounded from above by
\be\label{v22}
v_2^{pp} (p_T) \, \le \, \frac{\int d^2 k \, \frac{d
N_{corr}^{pp}}{d^3 p \, d^3 k} \, \cos 2 (\phi_p - \phi_k)}{\frac{d
N^{pp}}{d^3 p} \,
\frac{d N^{pp}}{d y_k}} \, \sqrt{\frac{\frac{d N^{pp}}{d y_p} 
\, \frac{d N^{pp}}{d y_k} }{\int d^2 p \,
d^2 k \frac{d N_{corr}^{pp}}{d^3 p \, d^3 k} \, \cos 2 (\phi_p -
\phi_k)}}.
\ee
We want to estimate the contribution to $v_2^{AA} (p_T)$ of the
non-flow correlations of the same physical origin as the ones giving
rise to $v_2^{pp} (p_T)$ in \eq{v21}. The contribution is
\be\label{v2AA}
v_2^{AA} (p_T) |_{non-flow} \, = \, \frac{\int d^2 k \, \frac{d
N_{corr}^{AA}}{d^3 p \, d^3 k} \, \cos 2 (\phi_p - \phi_k)}{\frac{d
N^{AA}}{d^3 p} \,
\frac{d N^{AA}}{d y_k}} \, \sqrt{\frac{\frac{d N^{AA}}{d y_p} 
\, \frac{d N^{AA}}{d y_k} }{\int d^2 p \,
d^2 k \frac{d N_{corr}^{AA}}{d^3 p \, d^3 k} \, \cos 2 (\phi_p -
\phi_k)}}.
\ee
Using the fact that, approximately, both in saturation models and in
the data $dN/dy \sim N_{part}$ \cite{KN,KL}, we rewrite the right hand
side of
\eq{v22} as
\be\label{v23}
v_2^{pp} (p_T) \, \le \, v_2^{AA} (p_T)|_{non-flow} \
\sqrt{\frac{N^{AA}}{N^{pp}}},
\ee
with $N^{AA}$ and $N^{pp}$ total particle multiplicities in $AA$ and
$pp$ collisions, proportional to the average number of participants
involved. Inverting \eq{v23} we obtain a lower bound 
\be\label{bound}
v_2^{AA} (p_T)|_{non-flow} \, \ge \, v_2^{pp} (p_T) \,
\sqrt{\frac{N^{pp}}{N^{AA}}}.
\ee
Preliminary analysis of $pp$ data yields $v_2^{pp} (p_T) \approx 1$ at
high-$p_T$ \cite{rs}. To get a lower bound we use $N^{AA}_{part} =
394$ and $N^{pp}_{part} = 2$ in \eq{bound} obtaining $v_2^{AA}
(p_T)|_{non-flow} \gsim 7 \%$.


\begin{theacknowledgments}
This work was supported in part by the U.S. Department of Energy under
Grant No. DE-FG03-97ER41014. The preprint number is NT@UW-03-020.
\end{theacknowledgments}


\bibliographystyle{aipproc}   

\end{document}